%% file: causalTesting.tex
\newcommand{\ourtitle}{Causal Testing: Understanding Defects' Root Causes}
\newcommand{\pdftitle}{Causal Testing: Understanding Defects' Root Causes}
\newcommand{\pdfauthors}{Brittan Johnson, Yuriy Brun, Alexandra Meliou}

\documentclass[sigconf, screen]{acmart}

\input{preamble}

\usepackage{listings}

\copyrightyear{2020}
\acmYear{2020}
\setcopyright{acmlicensed}
\acmConference[ICSE '20]{42nd International Conference on Software Engineering}{May 23--29, 2020}{Seoul, Republic of Korea}
\acmBooktitle{42nd International Conference on Software Engineering (ICSE '20), May 23--29, 2020, Seoul, Republic of Korea}
\acmPrice{15.00}
\acmDOI{10.1145/3377811.3380377}
\acmISBN{978-1-4503-7121-6/20/05}

\begin{document}



\title{\ourtitle}

\author{Brittany Johnson}
\affiliation{%
	\institution{University of Massachusetts Amherst}
	\city{Amherst}
	\state{MA}
    \postcode{01003-9264}
	\country{USA}}
\email{bjohnson@cs.umass.edu}

\author{Yuriy Brun}
\orcid{0000-0003-3027-7986}
\affiliation{%
	\institution{University of Massachusetts Amherst}
	\city{Amherst}
	\state{MA}
    \postcode{01003-9264}
	\country{USA}}
\email{brun@cs.umass.edu}

\author{Alexandra Meliou}
\affiliation{%
	\institution{University of Massachusetts Amherst}
	\city{Amherst}
	\state{MA}
    \postcode{01003-9264}
	\country{USA}}
\email{ameli@cs.umass.edu}

\begin{abstract}

Understanding the root cause of a defect is critical to isolating and
repairing buggy behavior. We present Causal Testing, a new method of
root-cause analysis that relies on the theory of counterfactual causality to
identify a set of executions that likely hold key causal information
necessary to understand and repair buggy behavior. Using the Defects4J
benchmark, we find that Causal Testing could be applied to 71\% of real-world
defects, and for 77\% of those, it can help developers identify the root
cause of the defect. A controlled experiment with 37 developers shows that
Causal Testing improves participants' ability to identify the cause of the
defect from 80\% of the time with standard testing tools to 86\% of the time
with Causal Testing. The participants report that Causal Testing provides
useful information they cannot get using tools such as JUnit. Holmes, our
prototype, open-source Eclipse plugin implementation of Causal Testing, is
available at \url{http://holmes.cs.umass.edu/}.

\end{abstract}

\maketitle

\section{Introduction}
\label{sec:Introduction}

Debugging and understanding software behavior is an important part of
building software systems. To help developers debug, many existing
approaches, such as spectrum-based fault localization~\cite{Jones02,
Souza16}, aim to automatically localize bugs to a specific location in the
code~\cite{Artzi10, Campos13}. However, finding the relevant line is often
not enough to help fix the bug~\cite{Parnin11}. Instead, developers need help
identifying and understanding the root cause of buggy behavior. While
techniques such as delta debugging can minimize a failing test
input~\cite{Zeller02} and a set of test-breaking changes~\cite{Zeller99},
they do not help explain \emph{why} the code is faulty~\cite{Johnson13}.

To address this shortcoming of modern debugging tools, this paper presents
\emph{Causal Testing}, a novel technique for identifying root causes of
failing executions based on the theory of counterfactual causality. Causal
Testing takes a manipulationist approach to causal
inference~\cite{Woodward05}, modifying and executing tests to observe causal
relationships and derive causal claims about the defects' root causes.

Given one or more failing executions, Causal Testing conducts \emph{causal
experiments} by modifying the existing tests to produce a small set of
executions that differ minimally from the failing ones but do not exhibit the
faulty behavior. By observing a behavior and then purposefully changing the
input to observe the behavioral changes, Causal Testing infers causal
relationships~\cite{Woodward05}: The change in the input \emph{causes} the
behavioral change. Causal Testing looks for two kinds of minimally-different
executions, ones whose inputs are similar and ones whose execution paths are
similar. When the differences between executions, either in the inputs or in
the execution paths, are small, but exhibit different test behavior, these
small, causal differences can help developers understand what is causing the
faulty behavior.

Consider a developer working on a web-based geo-mapping service (such as
Google Maps or MapQuest) receiving a bug report that the directions
between ``New York, NY, USA'' and ``900 Ren\'e L\'evesque Blvd.\ W Montreal,
QC, Canada'' are wrong.
The developer replicates the faulty
behavior and hypothesizes potential causes. Maybe the special characters in
``Ren\'e L\'evesque'' caused a problem. Maybe the first address being a city
and the second a specific building caused a mismatch in internal data types.
Maybe the route is too long and the service's precomputing of some routes is
causing the problem. Maybe construction on the Tappan Zee Bridge along the
route has created flawed route information in the database. There are many
possible causes to consider. The developer decides to step through
the faulty execution, but the shortest path algorithm coupled with
precomputed-route caching and many optimizations is complex, and it is not
clear how the wrong route is produced. The developer gets lost inside the
many libraries and cache calls, and the stack trace quickly becomes unmanageable. 

\begin{figure}[t]
\begin{lstlisting}
<@\textcolor{red}{Failing:}@> New York, NY, USA to                                900 René Lévesque Blvd. W Montreal, QC, Canada
<@\textcolor{red}{Failing:}@> Boston, MA, USA to                                  900 René Lévesque Blvd. W Montreal, QC, Canada
<@\textcolor{red}{Failing:}@> New York, NY, USA to                                  1 Harbour Square, Toronto, ON, Canada
<@\textcolor{drkgreen}{Passing:}@> New York, NY, USA to                                 39 Dalton St, Boston, MA, USA
<@\textcolor{drkgreen}{Passing:}@> Toronto, ON, Canada to                              900 René Lévesque Blvd. W Montreal, QC, Canada
<@\textcolor{drkgreen}{Passing:}@> Vancouver, BC, Canada to                            900 René Lévesque Blvd. W Montreal, QC, Canada
\end{lstlisting}
\vspace{-0.5ex}
\hrule
\smallskip
\lt{Minimally-different execution traces:}
\begin{lstlisting}[firstnumber=7]
<@\textcolor{red}{Failing:}@>                          <@\textcolor{drkgreen}{Passing:}@>
[...]                            [...]
findSubEndPoints(sor6, tar7);    findSubEndPoints(sor6, tar7);
findSubEndPoints(sor7, tar8);    findSubEndPoints(sor7, tar8);
<@\textcolor{red}{metricConvert(pathSoFar)}@>;
findSubEndPoints(sor8, tar9);    findSubEndPoints(sor8, tar9);
[...]                            [...]
\end{lstlisting}
\caption{Passing and failing tests for a geo-mapping service application, and
test execution traces.}
\label{fig:exampleTests}
\end{figure}

\looseness-1
Suppose, instead, a tool had analyzed the bug report's test and presented the
developer with the information in Figure~\ref{fig:exampleTests}.
The developer would quickly see that the special characters, the first
address being a city, the length of the route, and the construction are not
the root cause of the problem. Instead, all the failing test cases have one
address in the United States and the other in Canada, whereas all the passing
test cases have both the starting and ending addresses in the same country.
Further, the tool found a passing and a failing input with minimal execution
trace differences: the failing execution contains a call to the
\lt{metricConvert(pathSoFar)} method but the passing one does not.\footnote{Note that
prior work, such as spectrum-based fault localization~\cite{Jones02,
Souza16}, can identify the differences in the traces of existing tests; the
key contribution of the tool we describe here is generating the relevant
executions with the goal of minimizing input and execution trace
differences.} Armed with this information, the developer is now better
equipped to find and edit code to address the root cause of the bug.

We implement Causal Testing in an open-source, proof-of-con\-cept Eclipse
plug-in, Holmes, that works on Java programs and interfaces with JUnit.
Holmes is publicly available at \url{http://holmes.cs.umass.edu/}.
We evaluate Causal Testing in two ways. First, we use Holmes in a
controlled experiment. We asked 37 developers to identify the root causes of
real-world defects, with and without access to Holmes. We found that
developers could identify the root cause 86\% of the time when using Holmes,
but only 80\% of the time without it.
Second, we evaluate Causal Testing's applicability to real-world defects by
considering defects from real-world programs in the Defects4J
benchmark~\cite{Just14defects4j}. We found that Causal Testing could be
applied to 71\% of real-world defects, and that for 77\% of those, it could
help developers identify the root cause.

\begin{figure*}[t]
  \centering
  \includegraphics[width=\textwidth]{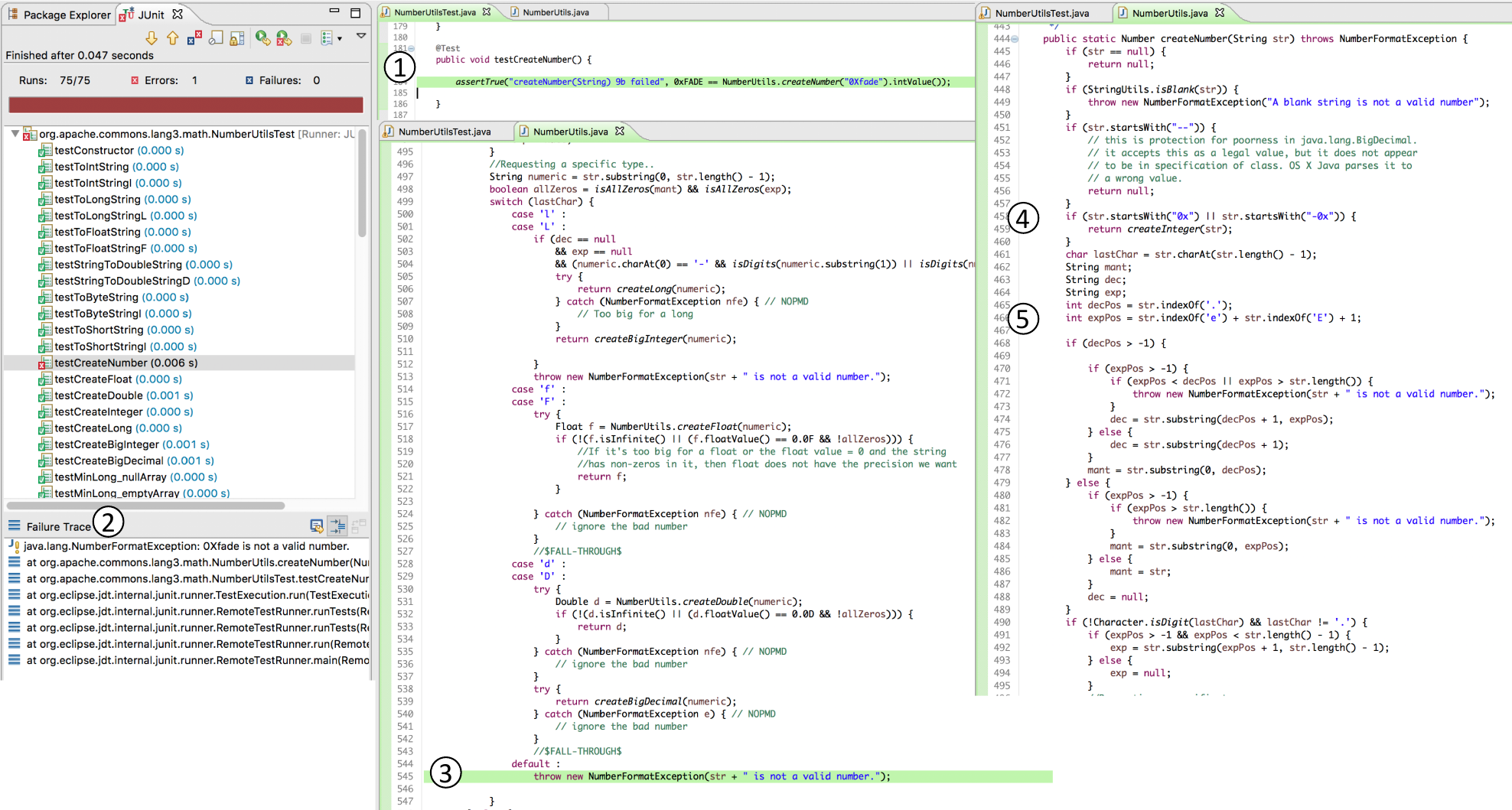}
  \caption{Amaya's Eclipse IDE, while she is debugging a defect evidenced by
  a failing test.}
	\label{fig:code}
\end{figure*}

A rich body of prior research aims to help developers debug faulty behavior.
Earlier-mentioned fault localization techniques~\cite{Agrawal95, Artzi10,
Campos13, Hao05, Hao05a, Jones02, Menzies07, Menzies10, Souza16, Wong10,
Zuddas14} rank code locations according to the likelihood that they contain a
fault, for example using test cases~\cite{Jones02} or static code
features~\cite{Menzies07, Menzies10}. The test-based rankings can be
improved, for example, by generating extra tests~\cite{Artzi10, Zuddas14} or
by applying statistical causal inference to observational data~\cite{Baah10,
Baah11}. Automated test generation can create new tests, which can help
discover buggy behavior and debug it~\cite{Godefroid07, Godefroid08,
Holler12, Jung08}, and techniques can minimize test suites~\cite{Jin10,
Orso04, Wang18a} and individual tests~\cite{Zeller99, Zeller02,
Hildebrandt00} to help deliver the most relevant debugging information to the
developer. These techniques can help developers identify \emph{where} the bug
is. By contrast, Causal Testing focuses on explaining \emph{why} buggy
behavior is taking place. Unlike these prior techniques, Causal Testing
generates \emph{pairs of very similar tests} that nonetheless exhibit
different behavior. Relatedly, considering tests that exhibit minimally
different behavior, BugEx focuses on tests that differ slightly in branching
behavior~\cite{Rosler12} and Darwin generates tests that pass a version of
the program without the defect but fail a version with the
defect~\cite{Qi12}. Unlike these techniques, Causal Testing requires only a
single, faulty version of the code, and only a single failing test, and then
conducts causal experiments and uses the theory of counterfactual causality
to produce minimally-different tests and executions that help developers
\emph{understand} the cause of the underlying defect.

The rest of this paper is structured as follows. 
Section~\ref{sec:motivating} illustrates how Causal Testing can help
developers on a real-world defect.
Sections~\ref{sec:approach}~and~\ref{sec:holmes} describe Causal Testing and
Holmes, respectively.
Section~\ref{sec:evalUtility} evaluates how useful Holmes is in identifying
root causes and Section~\ref{sec:ApplicabilityEvaluation} evaluates how
applicable Causal Testing is to real-world defects.
Section~\ref{sec:Discussion} discusses the implications of our findings and
limitations and threats to the validity of our work.
Finally Section~\ref{sec:related} places our work in the context of related
research, and Section~\ref{sec:Contributions} summarizes our contributions.

\section{Motivating Example}
\label{sec:motivating}

Consider Amaya, a developer who regularly contributes to open source
projects. Amaya codes primarily in Java and regularly uses the Eclipse IDE
and JUnit. Amaya is working on addressing a bug report in the Apache Commons
Lang project. The report comes with a failing test (see \circled{1} in
Figure~\ref{fig:code}).

\looseness-1
Figure~\ref{fig:code} shows Amaya's IDE as she works on this bug. Amaya runs
the test to reproduce the error and JUnit reports that an exception occurred
while trying to create the number \lt{0Xfade} (see \circled{2} in
Figure~\ref{fig:code}).
Amaya looks through the JUnit failure trace, looking for the place the code
threw the exception 
(see \circled{3} Figure~\ref{fig:code}). 
Amaya observes that the exception comes from within a \lt{switch} statment,
and that there is no case for the \lt{e} at the end of \lt{0Xfade}. To add
such a case, Amaya examines the other \lt{switch} cases and realizes that
each case is making a different kind of number, e.g., the case for \lt{l}
creates either a \lt{long} or \lt{BigInteger}. Since \lt{0Xfade} is 64222,
Amaya conjectures that this number fits in an \lt{int}, and creates a new
method call to \lt{createInteger()} inside of the case for \lt{e}.
Unfortunately, the test still fails.

Using the debugger to step through the test's execution, Amaya sees the
\lt{NumberFormatException} thrown on line~545 (see \circled{3} in
Figure~\ref{fig:code}). She sees that there are two other locations the input
touches (see \circled{4} and \circled{5} in Figure~\ref{fig:code}) during
execution that could be affecting the outcome. She now realizes that the code
on lines~497--545, despite being where the exception was thrown, may not be
the location of the defect's cause. She is feeling stuck.

But then, Amaya remembers a friend telling her about Holmes, a Causal Testing
Eclipse plug-in that helps developers debug. Holmes tells her that the code
fails on the input \lt{0Xfade}, but passes on input \lt{0xfade}. The key
difference is the lower case \lt{x}. Also, according to the execution trace
provided by Holmes, these inputs differ in the execution of line~458 (see
\circled{4} in Figure~\ref{fig:code}). The \lt{if} statement
fails to check for the \lt{0X} prefix. Now, armed with the cause of the
defect, Amaya turns to the Internet to find out the hexadecimal specification
and learns that the test is right, \lt{0X} and \lt{0x} are both valid
prefixes for hexadecimal numbers. She augments the \lt{if} statement and the
bug is resolved!

Holmes implements Causal Testing, a new technique for helping understand root
causes of behavior. Holmes takes a failing test case (or test cases) and
perturbs its inputs to generate a pool of possible inputs. For example,
Holmes may perturb \lt{0Xfade} to \lt{0XFADE}, \lt{0xfade}, \lt{edafX},
\lt{0Xfad}, \lt{Xfade}, \lt{fade}, and many more. Holmes then executes all
these inputs to find those that pass the original test's oracle, and, next,
selects from the passing test cases a small number such that either their inputs or
their execution traces are the most similar to the original, failing test
case. Those most-similar passing test cases help the
developer understand the key input difference that makes the test pass.
Sometimes, Holmes may find other failing test cases whose inputs are even
more similar to the passing ones than the original input, and it would report
those too. The idea is to show the smallest difference that causes the
behavior to change. 

Holmes presents both the static (test input) and dynamic (execution trace)
information to the developer to compare the minimally-different passing and
failing executions to better understand the root cause of the bug. For
example, for this bug, Holmes shows the inputs, \lt{0Xfade} and
\lt{0xfade}, and the traces of the two executions, showing that the passing 
test enters a method 
from \lt{createInteger} that the failing test cases do not, dictating
to Amaya the expected code behavior, leading her to fix the bug.

\section{Causal Testing}
\label{sec:approach}

Amaya's debugging experience is based on what actual developers did while
debugging real defects in a real-world version of Apache Commons Lang (taken
from the Defects4J benchmark~\cite{Just14defects4j}). As the example
illustrates, software is complex and identifying root causes of
program failures is challenging. This section describes our Causal Testing
approach to computing and presenting developers with information that can
help identify root causes of failures.

\begin{figure}[b]
	\includegraphics[width=\columnwidth]{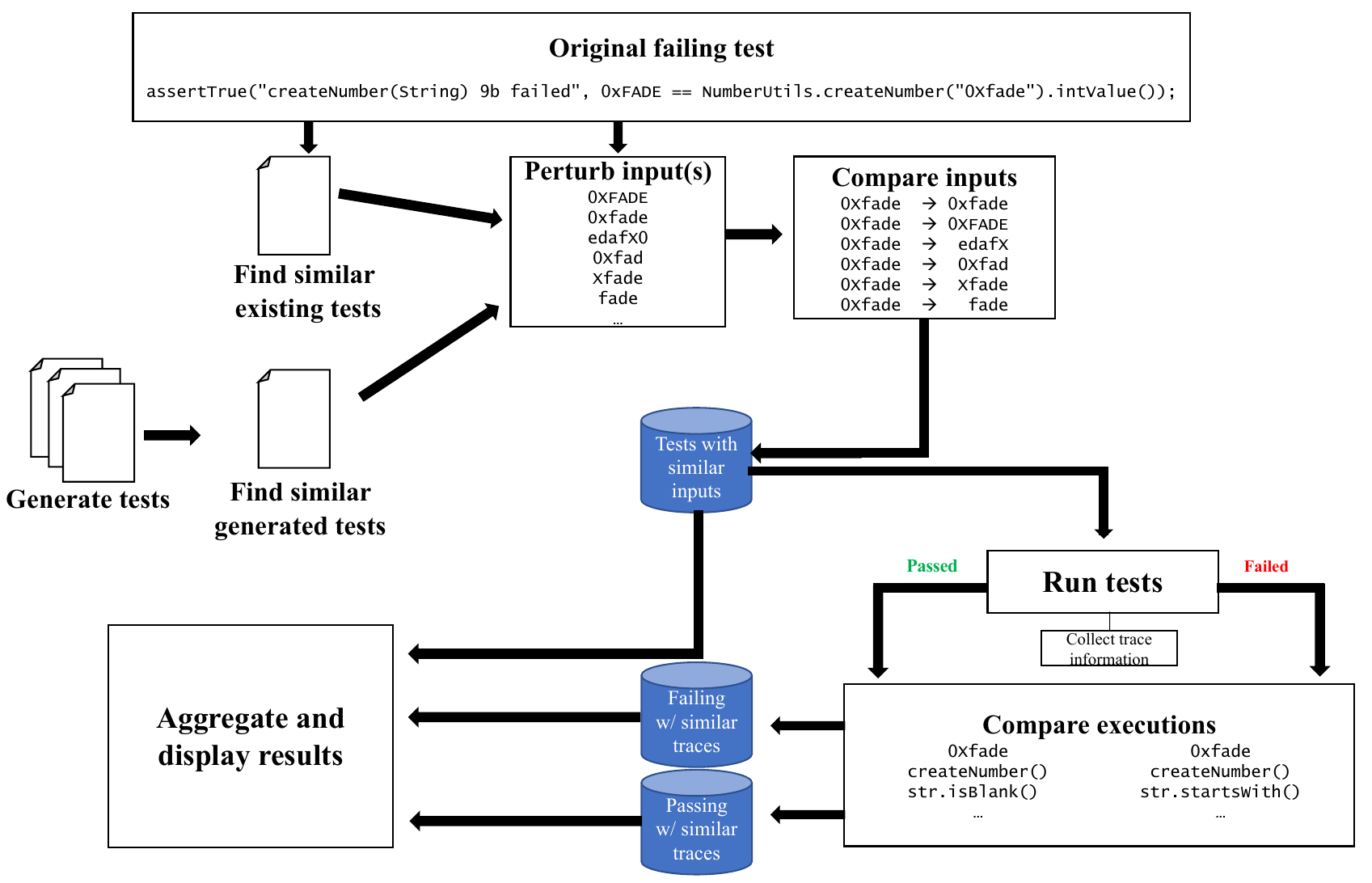}
	\caption{Causal Testing computes minimally-different test inputs that,
		nevertheless, produce different behavior.}
	\label{fig:causal}
\end{figure}

Figure~\ref{fig:causal} describes the Causal Testing approach. Given a
failing test, Causal Testing conducts a series of causal experiments starting
with the original test suite. Causal Testing provides experimental results to
developers in the form of minimally-different passing and failing tests, and
traces of their executions.

\subsection{Causal Experiments with Test Cases}

Causal Testing modifies test cases to conduct causal experiments; it observes
system behavior and then reports the changes to test inputs that cause system
behavior to change. To create these test case modifications and to then
identify the modifications that lead to behavioral change, Causal Testing
needs a systematic way of perturbing inputs and of measuring test case
similarity, which we describe in this section. Once the experiments are
complete, Causal Testing reports to the developer a list of minimally
different passing and failing test case inputs and their execution traces, to
help explain root causes of the failing behavior.

\subsubsection{Perturbing Test Inputs}
\label{subsec:fuzzing}

To conduct causal experiments, Causal Testing starts with a failing test,
which we shall call from now on the \emph{original failing test}, and
identifies the class this test is testing. Causal Testing considers all the
tests of that class, and generates more tests using automated test input
generation (and the oracle from the one failing test), to create a set of
failing and passing tests. Then, Causal Testing fuzzes these existing and
generated test inputs to find additional tests that exhibit expected and
unexpected behavior.

Theoretically, it is also possible for Causal Testing to perturb the test
oracle. For example, it might change the \lt{assertTrue} in
Figure~\ref{fig:causal} to \lt{assertFalse}. However, perturbing test oracles
is unlikely to produce meaningful information to guide the developer to the
root cause, or, at least, is likely to produce misleading information. For
example, making a test pass simply by changing the oracle does not provide
information about key differences in test \emph{inputs} that alter software
behavior. As such, Causal Testing focuses on perturbing test inputs only.

There are different ways Causal Testing could assemble sets of passing and
failing tests. First, Causal Testing could simply rely on the tests already
in the test suite. Second, Causal Testing could use automated test
generation~\cite{afl, Fraser13, Pacheco07a} to generate a large number of
test inputs. Third, Causal Testing could use test fuzzing to change the
existing tests' inputs to generate new, similar inputs. Fuzz testing is an
active research area~\cite{Godefroid07, Godefroid08, Holler12, Jung08,
Walls15issre} (although the term fuzz testing is also used to mean simply
generating tests~\cite{afl}) and has been applied in the security domain to
stress-test an application and automatically discover vulnerabilities,
e.g.,~\cite{Godefroid08, Holler12, Walls15issre}.

While in real-world systems, existing test suites often contain both passing
and failing tests, these suites are unlikely to have similar enough pairs of
one passing, one failing tests to provide useful information about the root
cause. Still, it is worthwhile to consider these tests first, before trying
to generate more. As such, our solution to the challenge of generating
similar inputs is to (1)~start with all existing tests, (2)~use multiple
fuzzers to fuzz these tests, (3)~generate many tests, and (4)~filter those
tests to select the ones similar to the original failing test. As we observed
with Holmes, our proof-of-concept Causal Testing tool (described in
Section~\ref{sec:holmes}), using multiple input fuzzers provided a diverse
set of perturbations, increasing the chances that Causal Testing finds a set
of minimally-different inputs and that at least one of them would lead to a
passing execution.

\subsubsection{Input Similarity}
\label{subsec:comparison}

Given two tests that differ in their inputs but share the same oracle, Causal
Testing needs to measure the similarity between the two tests, as its goal is
to find pairs of minimally-different tests that exhibit opposite behavior.
Conceptually, to apply the theory of causal inference, the two tests should
differ in only one factor. For example, imagine a software system that
processes apartment rental applications. If two application inputs are
identical in every way except one entry, and the software crashes on one but
not on the other, this pair of inputs provides one piece of evidence that the
differing entry \emph{causes} the software to crash. (Other pairs that also
only differ in that one entry would provide more such evidence.) If the
inputs differed in multiple entries, it would be harder to know which entry
is responsible. Thus, to help developers understand root causes, Causal
Testing needs to precisely measure input similarity. We propose two ways to
measure input similarity: \emph{syntactic differences} and \emph{execution
path differences}.

\textbf{Static Input Differences.}
The static input similarity can be viewed at different scopes. First, inputs
can agree in some and differ in others of their arguments (e.g., parameters
of a method call). Agreement across more arguments makes inputs more similar.
Second, each argument whose values for the two tests differ can differ to
varying degrees. A measure of that difference depends on the type of the
argument. For arguments of type \lt{String}, the Levenshtein distance (the
minimum number of single-character edits required to change one \lt{String}
into the other) is a reasonable measure, though there are others as well,
such as Hamming distance (difference between two values at the bit level).
For numerical arguments, their numerical difference or ratio is often a
reasonable measure.

We found that relatively simple measures of similarity suffice for general
debugging, and likely work well in many domains. Using Levenshtein or Hamming
distance for \lt{String}s, the arithmetic difference for numerical values,
and sums of elements distances for \lt{Array}s, worked reasonably well, in
practice, on the 330 defects from four different real-world systems we
examined from the Defects4J benchmark~\cite{Just14defects4j}. However, more
generally, the semantics of similarity measures are dependent on the domain.
Some arguments may play a bigger role than others, and the meaning of some
types may only make sense in the particular domain. For example, in apartment
rental applications, a difference in the address may play a much smaller role
than a difference in salary or credit history. As such, how the similarity of
each argument is measured, and how the similarities of the different
arguments are weighed are specific to the domain and may require fine tuning
by the developer, especially for custom data types (e.g., project-specific
\lt{Object} types). Still, in the end, we found that simple, domain-agnostic
measures worked well in the domains we examined.  

\textbf{Execution Path Differences.}
Along with static differences, two inputs can differ based on their dynamic
behavior at runtime. One challenge when considering only static input
differences is that a statically similar input may not always yield an
outcome that is relevant to the original execution. For example, it is
possible that two inputs that differ in only one character lead to completely
incomparable, unrelated executions. Therefore, Causal Testing also collects
and compares dynamic information in the form of the execution path the input
causes.

Beyond simplistic ways to compare executions, such as by their lengths,
comparing the statements and method calls in each execution provides
information we found helpful to understanding root causes. This also
strengthens the causal connection between the input change and the behavior
change; if two inputs' executions, one passing and one failing, only differ
by one executed statement, it is likely that one statement plays an important
role in the behavioral change. Augmenting method calls with their return
values provides additional insights in situations where the bug is evident
not by the sequence of statements executed but in the use of a method that
returns an unexpected value.

\bigskip 
Both static and execution path measures of similarity can be useful in
identifying relevant tests that convey useful information to developers.
Inputs that are similar both statically and in terms of execution paths hold
potential to convey even more useful information, as they have even fewer
differences with the original failing test. Therefore, Causal Testing
prioritizes tests whose inputs are statically and dynamically similar to the
original failing test.

\subsection{Communicating Root Causes to Developers}
\label{subsec:test_results}

After generating and executing test inputs, Causal Testing ranks them by
similarity and selects a user-specified target number of the most similar
passing test cases. In our experience, three tests was a good target, though,
at times, a time-out was necessary because finding three similar passing
tests was computationally infeasible. Causal Testing reports tests as it
finds them, produce results for the developer as quickly as possible, while
it performs more computation, looking for potentially more results.

Causal Testing collects the input and the execution traces for each test it
executes. These are, of course, used for determining test case similarity,
but also hold the key information in terms of what differences in test inputs
lead to what behavioral changes. For the pairs of failing and passing tests,
Causal Testing presents the static differences in inputs, and the execution
traces (along with each method call's arguments and return values) with
differences highlighted. Because execution traces can get large, parsing them
can be difficult for developers; showing differences in the traces simplifies
this task. Causal Testing displays a minimized trace, focused on the
differences.

\section{Holmes: A Causal Testing Prototype}
\label{sec:holmes}

We have implemented Holmes, an open source Eclipse plug-in Causal Testing
prototype. Holmes is available at \url{http://holmes.cs.umass.edu/}
and consists of four components: \emph{input and test case generators},
\emph{edit distance calculators \& comparers}, a \emph{test executor \&
comparator}, and an \emph{output view}.

\subsection{Input \& Test Case Generation}

Holmes first task is to create a set of candidate test cases. Holmes first
searches all tests in the current test suite for tests that are similar to
the original failing test using string matching to determine if two tests are
similar. More specifically, Holmes converts the entire test file to a string
and parses it line by line. This is an approximation of test similarity.
Future work can improve Holmes by considering similarity in dynamic execution
information between the two tests, or by creating new tests by using test
inputs from other tests but the oracle from the original failing test.

Next, Holmes proceeds to generate new tests. Holmes gets new inputs for
generating new tests in two ways:

\begin{itemize}

  \item \textbf{Test case generation.} Holmes uses an existing test case
  generation tool, EvoSuite~\cite{Fraser13}. We chose EvoSuite because it is
  a state-of-the-art, open-source tool that works with Java and JUnit. Holmes
  determines the target class to generate tests from based on the class the
  original failing test tests. For example, if the original failing test is
  called \mbox{\lt{NumberUtilsTest},} Holmes tells EvoSuite to generate tests
  for \lt{NumberUtils}. To determine if a test is related to the original
  failure, Holmes searches the generated tests for test cases that call the
  same method as the original test. From this process, Holmes will get at
  least one valid input to use during fuzzing.
	
  \item \textbf{Input fuzzing.} To generate additional inputs for new tests,
  Holmes fuzzes existing and generated test inputs. Holmes uses two
  off-the-shelf, open-source fuzzers,
  \textsc{Peach}\footnote{\url{https://github.com/MozillaSecurity/peach}} and
  \textsc{Fuzzer}\footnote{\url{https://github.com/mapbox/fuzzer }}. To
  increase the chances that fuzzed inputs will produce passing tests, Holmes
  prioritizes (when available) inputs from passing tests. Holmes fuzzes the
  original input and all valid inputs from generated test cases, again to
  increase the chance of finding passing tests.
	
\end{itemize}

Once Holmes runs test generation and fuzzes the valid inputs, the next step
is to determine which of the generated inputs are most similar to the
original.

\subsection{Test Execution \& Edit Distance Calculation}

The current Holmes implementation uses static input similarity to identify
minimally-different tests. Using only static input similarity first provided
us with a better understanding of how execution information could be
collected and used most effectively. In the user study described in
Section~\ref{sec:evalUtility}, we semi-automated using dynamic execution
trace information for evaluating Holmes. Future work can improve Holmes by
automatically using dynamic execution trace information, as described in
Section~\ref{subsec:comparison}.

To evaluate static input differences, Holmes first determines the data type
of each argument in the method-under-test; this determines how Holmes will
calculate edit distance. For arguments with numerical values, Holmes
calculates the absolute value of the arithmetic difference between the
original and generated test input argument. For example, inputs 1.0 and 4.0
have an edit distance of 3.0. For \lt{String} and \lt{char} inputs, Holmes
uses two different metrics. First, Holmes determines the Hamming distance
between the two arguments. We elected to use Hamming distance first because
we found it increases the accuracy of the similarity measure for randomly
generated inputs. Once Holmes identifies inputs that are similar using the
Hamming distance, it uses the Levenshtein distance to further refine its
findings; inputs that require the fewest character changes to change from one
to the other are most similar. Holmes uses an edit distance threshold of 3;
tests whose inputs are more than a Levenshtein distance of 3 away from the
original failing tests are considered too different to be reported to the
developer.

Holmes uses the executed test behavior to determine which inputs satisfy the
original failing test's oracle. Then, Holmes attempts to further minimize the
test differences by, for each original argument, iteratively replacing the
original value with new input value and executing the modified test to
observe if the oracle is satisfied. Holmes iterates to try to find three
similar passing tests to compare to the failing one.

\subsection{Communicating Root Causes to Developers}

An important consideration when building a tool is how it will communicate
with the developer~\cite{Johnson16}. Once Holmes has computed a set of
passing (and a set of failing) tests, it organizes the information for
presentation. Holmes organizes tests by whether it passes or fails, showing
the original failing test at the top of the output window, making it easy to
compare the differences. Under each test, Holmes presents a minimized test
execution trace. So as to not overwhelm the developer with information,
Holmes' user interface includes the option to toggle showing and hiding trace
information.

\subsection{Holmes' Limitations}

We implemented Holmes as a prototype Causal Testing tool, to be used in a
controlled experiment with real users (see Section~\ref{sec:evalUtility}). We
have thus prioritized ensuring Holmes implements the aspects of Causal
Testing we needed to evaluate, over fully automating it.

The current version of Holmes automates test generation, execution, and
static edit distance calculation. We used InTrace~\cite{InTrace} to collect
runtime execution traces and then \emph{manually} incorporated the execution
information with the tests. Future versions of Holmes will automate the
dynamic trace collection and comparison.

The current version of Holmes relies on the Defects4J
benchmark~\cite{Just14defects4j} used in our evaluations, and extending it to
other defects may require extending Holmes or setting those defects' projects
up in a particular way. For simplicity, Holmes works on single-argument tests
with \lt{String} or primitive arguments. While this is sufficient for the
defects in Defects4J benchmark, this limitation will need to be lifted for
tests with multiple arguments. Our Holmes prototype implementation is
open-source, to allow others to build on it and improve it.

\section{Causal Testing Effectiveness}
\label{sec:evalUtility}

We designed a controlled user study experiment with 37~developers to answer
the following three research questions:

\begin{itemize}
	
	\item[RQ1:] Does Causal Testing improve the developers'
	ability to identify the root causes of defects?
	
	\item[RQ2:] Does Causal Testing improve the developers'
	ability to repair defects?
	
	\item[RQ3:] Do developers find Causal Testing useful, and, if
	so, what aspect of Causal Testing is most useful?
	
\end{itemize}

\subsection{User Study Design}
\label{subsec:design}

Causal Testing's goal is to help developers determine the cause of a test
failure, thereby helping developers better understand and eliminate defects
from their code. We designed our user study and prototype version of Holmes
to provide evidence of Causal Testing's usefulness, while also providing a
foundation of what information is useful for Causal Testing.

We randomly selected seven defects from Defects4J, from the Apache Commons
Lang project. We chose Apache Commons Lang because it (1)~is the most widely
known project in Defects4J, (2)~had defects that required only limited domain
knowledge, and (3)~can be developed in Eclipse.

Our user study consisted of a training task and six experimental tasks. Each
task mapped to one of the seven defects. Each participant started with the
training task, and then performed six experimental tasks. The training task
and three of the experimental tasks used Holmes and the other three
experimental tasks belonged to the control group and did not include the use
of Holmes. The order of the tasks, and which tasks were part of the control
group and which part of the experimental group were all randomized.

For the training task, we provided an Eclipse project with a defective code
version and single failing test. We explained how to execute the test suite
via JUnit, and how to invoke Holmes. We allowed participants to explore the
code and ask questions, telling them that the goal is to change the code so
that all that tests pass. Each task that followed was similar to the training
task; control group tasks did not have access to Holmes, experimental group
tasks did.

We recorded audio and the screen for later analysis. We asked participants to
complete a causality questionnaire after each task consisting of two
questions: ``What caused Test X to fail?'' and ``What changes did you make
to fix it?''

At then end, the participants completed an exit survey with open-ended
questions, such as ``What information did you find most helpful when
determining what caused tests to fail?'' and 4-point Likert scale questions,
such as ``How useful did you find X?'' For the Likert-scale questions, we
gave participants the options ``very useful'', ``somewhat useful'', ``not
useful'', and ``misleading or harmful''. We also gave participants an
opportunity to provide additional feedback they saw fit.

Prior to our experiment, we conducted a pilot of our initial user study
design with 23~students from a graduate software engineering course. Our
pilot study consisted of 5~tasks and a mock-up version of Holmes. We used
lessons learned and challenges encountered to finalize the design of our
study. The 23~pilot participants did not participate in the final study
presented here. All final study materials are available online at
\url{http://holmes.cs.umass.edu} in the \texttt{user\_study\_materials}
directory.

\subsection{Participants}
\label{subsec:participants}

We recruited a total of 39~participants from industry and academia:
15~undergraduate students, 12~PhD students, 9~Masters students, 2~industry
developers, and 1~research scientist. Participants' programming experience
ranged from 1~to~30 years and experience with Java ranged from a few months
to 15~years. All participants reported having prior experience with Eclipse
and JUnit. We analyzed data from 37~participants; 2~undergraduate
participants (P2 and P3) did not follow the instructions, so we removed them from
our dataset.

\begin{figure}[t]
	\centering
	\small
	\begin{tabular}{ccccc}
		\toprule
		Defect &  Group                & Correct                 & Incorrect & Total\\
		\midrule
		\multirow{2}{*}{1} &  Control              & 17~(89\%)               & \phantom{0}2~(11\%)    & \phantom{0}19 \\
		&  Holmes               & 17~(94\%)    			& \phantom{0}1~\phantom{0}(5\%)      & \phantom{0}18  \\
		\midrule
		\multirow{2}{*}{2} &  Control              & 12~(60\%)    			& \phantom{0}8~(40\%)     & \phantom{0}20 \\
		&  Holmes               & \phantom{0}9~(53\%)    & \phantom{0}8~(47\%)    & \phantom{0}17 \\
		\midrule
		\multirow{2}{*}{3} &  Control              & 19~(95\%)    			& \phantom{0}1~\phantom{0}(5\%)   				& \phantom{0}20  \\
		&  Holmes               & 16~(94\%)    			& \phantom{0}1~\phantom{0}(6\%)   				& \phantom{0}17 \\
		\midrule
		\multirow{2}{*}{4} &  Control              & 15~(83\%)    			& \phantom{0}3~(17\%)    				& \phantom{0}18 \\
		&  Holmes               & 18~(95\%)    			& \phantom{0}1~\phantom{0}(5\%)    				& \phantom{0}19 \\
		\midrule
		\multirow{2}{*}{5} &  Control              & 13~(87\%)    			& \phantom{0}2~(13\%)    				& \phantom{0}15 \\
		&  Holmes               & 21~(95\%)    			& \phantom{0}1~\phantom{0}(5\%)    				& \phantom{0}22 \\
		\midrule
		\multirow{2}{*}{6} &  Control              & 12~(67\%)     			& \phantom{0}6~(33\%)    				& \phantom{0}18 \\
		&  Holmes               & 15~(79\%)     			& \phantom{0}4~(21\%)   				& \phantom{0}19 \\
		\midrule
		\midrule
		\multirow{2}{*}{Total} & Control           &\textbf{88 (80\%)}               & \textbf{22 (20\%)}   & 110             \\
		&  Holmes               & \textbf{96~(86\%)}              & \textbf{16~(14\%)}  & 112           \\
		\bottomrule
	\end{tabular}
	\caption{Distributions of correct and incorrect cause descriptions, per defect.}
	\label{fig:correctness}
\end{figure}

\subsection{User Study Findings}
\label{subsec:user_results}

We now summarize the results from our study.

\medskip
\noindent\textbf{RQ1: Does Causal Testing improve the developers' ability to
identify the root causes of defects?}

The primary goal of Causal Testing is to help developers identify the root
cause of test failures. To answer RQ1, we analyzed the responses participants
gave to the question ``What caused Test X to fail?'' We marked responses as
either correct (captured full and the true cause) or incorrect (missing part
or all of the true cause).

Figure~\ref{fig:correctness} shows the root cause identification correctness
results. When using Holmes, developers correctly identified the cause 86\% of
the time (96 out of 112 times). The control group only identified the cause
80\% of the time (88 out of 110). Fisher's exact test finds that these
samples come from different distributions with 83\% probability ($p = 0.17$).

For four of the six defects, (Defects~1,~4,~5,~and~6), developers using
Holmes were more accurate when identifying root causes than the control
group. For Defects~1,~4,~and~5, participants only incorrectly identified the
cause approximately 5\% of the time when using Holmes, compared to 11--17\%
of the time without Holmes. For Defect~6, participants with Holmes identified
the correct cause 79\% (15 out of 19) of the time; without Holmes they could
only identify the correct cause 67\% (12 out of 18) of the time. Our findings
suggest that \textbf{Causal Testing supports and improves developer ability
to understand root causes, for at least some defects.}

\medskip
\noindent\textbf{RQ2: Does Causal Testing improve the developers' ability to
repair defects?}

While Causal Testing's main goal is to help developers understand the root
cause, this understanding may be helpful in removing the defect as well. To
answer RQ2, we analyzed participants' responses to the question ``What
changes did you make to fix the code?'' We used the same evaluation criteria
and labeling as for RQ1. To determine if causal execution information
improves developers' ability to debug and repair defects, we observed the
time it took participants to complete each task and the correctness of their
repairs.

\begin{figure}[t]
	\centering
	\small
	\begin{tabular}{lccccccc}
		\toprule
		& \multicolumn{6}{c}{Average Resolution Time (in minutes)} \\
		\textbf{Defect:}   & \textbf{1} & \textbf{2} & \textbf{3} & \textbf{4} & \textbf{5} & \textbf{6} \\
		\midrule
		Control           & 16.5            & 10.6  		   & 6.8           & 12.9            & 3.7           & 10.0   \\
		Holmes            & 17.0            & 12.7            & 6.4          & 17.7            & 4.9           & 10.1	  \\
		\bottomrule
		
	\end{tabular}
	\caption{The average time developers took to resolve the defects, in minutes.}
	\label{fig:resolution_times}
\end{figure}

Figure~\ref{fig:resolution_times} shows the average time it took developers
to repair each defect. We omitted times for flawed repair attempts that do
not address the defect. On average, participants took more time with Holmes
on all but one defect (Defect~3). One explanation for this observation is
that while Holmes helps developers understand the root cause, this
understanding takes time, which can reduce the overall speed of repair.

\begin{figure}[t]
	\centering
	\small
	\begin{tabular}{ccccc}
		\toprule
		Defect &  Group                & Correct                 & Incorrect & Total\\
		\midrule
		\multirow{2}{*}{1} &  Control              & 16~\phantom{0}(89\%)               & \phantom{0}2~(11\%)    & 18 \\
		&  Holmes               & 12~\phantom{0}(86\%)    			& \phantom{0}2~(14\%)      & 14  \\
		\midrule
		\multirow{2}{*}{2} &  Control              & 12~(100\%)    			& \phantom{0}0~\phantom{0}(0\%)     & 12 \\
		&  Holmes               & \phantom{0}7~(100\%)    & \phantom{0}0~\phantom{0}(0\%)    & \phantom{0}7 \\
		\midrule
		\multirow{2}{*}{3} &  Control              & 19~(100\%)    			& \phantom{0}0~\phantom{0}(0\%)   				& 19  \\
		&  Holmes               & 16~(100\%)    			& \phantom{0}0~\phantom{0}(0\%)   				& 16 \\
		\midrule
		\multirow{2}{*}{4} &  Control              & 15~(100\%)    			& \phantom{0}0~\phantom{0}(0\%)    				& 15 \\
		&  Holmes               & 19~~(100\%)    			& \phantom{0}0~\phantom{0}(0\%)    				& 19 \\
		\midrule
		\multirow{2}{*}{5} &  Control              & 12~\phantom{0}(86\%)    			& \phantom{0}2~(14\%)    				& 14 \\
		&  Holmes               & 21~\phantom{0}(95\%)    			& \phantom{0}1~\phantom{0}(5\%)    				& 22 \\
		\midrule
		\multirow{2}{*}{6} &  Control              & \phantom{0}6~\phantom{0}(75\%)     			& \phantom{0}2~(25\%)    				& \phantom{0}8 \\
		&  Holmes               & \phantom{0}5~(100\%)     			& \phantom{0}0~\phantom{0}(0\%)   				& \phantom{0}5 \\
		\midrule
		\midrule
		\multirow{2}{*}{Total} & Control           & \textbf{80~\phantom{0}(93\%)}               & \phantom{0}\textbf{6~\phantom{0}(7\%)}   & 86             \\
		&  Holmes               & \textbf{80~\phantom{0}(96\%)}              & \phantom{0}\textbf{3~\phantom{0}(4\%)}  & 83           \\
		\bottomrule
	\end{tabular}
  \caption{Distribution of correct and incorrect repairs implemented by
  participants, per defect.}
	\label{fig:correct_fixes}
\end{figure}

Figure~\ref{fig:correct_fixes} shows repair correctness results. When using
Holmes, developers correctly repaired the defect 96\% of the time (80 out of
83) while the control group repaired the defect 93\% of the time (80 out of
86).

For two of the six defects (Defects~5~and~6), developers using Holmes
repaired the defect correctly more often (Defect~5: 95\%~vs.~86\%;
Defect~6:~100\%~vs.~75\%). For Defects~2,~3,~and~4, developers repaired the
defect correctly 100\% of the time both with and without Holmes. For one
defect (Defect~1), developers with Holmes were only able to repair the defect
correctly 86\% (12 out of 14) of the time while developers without Holmes
correctly fixed defects 100\% of the time.

Holmes did not demonstrate an observable advantage when repairing defects.
Our findings suggest that \textbf{Causal Testing sometimes helps developers
repair defects, but neither consistent\-ly nor statistically significantly.}

\medskip
\noindent\textbf{RQ3: Do developers find Causal Testing useful, and, if so,
what aspect of Causal Testing is most useful?}

To answer RQ3, we analyzed post-evaluation survey responses to the question
asking which information was most useful when understanding and debugging the
defects. We extracted and aggregated quantitative and qualitative results
regarding information most helpful when determining the cause of and fixing
the defects. We also analyzed the Likert-scale ratings regarding the
usefulness of JUnit and the various components of causal execution
information.

Overall, participants found the information provided by Holmes more useful
than other information available when understanding and debugging the
defects. Out of 37 participants, 17~(46\%) found the addition of at least one
aspect of Holmes more useful than output provided by JUnit alone. Further,
15~(41\%) participants found the addition of Holmes at least as useful as
JUnit. The remaining 5~(13\%) found the addition of Holmes not as useful as
JUnit alone. Though majority of participants found Holmes' output more
useful, JUnit and interactive debuggers are an important part of debugging.
Therefore, our expectations would be that Causal Testing would augment those
tools, not replace them.

Participants found the minimally-different passing tests Holmes provided the
most useful: 20 out of 37 participants~(54\%) rated this piece of information
as ``Very Useful.'' The passing and failing test inputs that Holmes provided
received ``Very Useful'' or ``Useful'' rankings more often than the test
execution traces. Finally, 18 participants marked either the passing or
failing execution trace as ``Not Useful.'' One participant felt the passing
test traces were ``Misleading or Harmful;'' during their session, they noted
that they felt in some cases the execution paths were not as similar as
others, which made interpreting the output more confusing.

To gain a better understanding of what parts of causal execution information
are most useful, and why, we also analyzed participants' qualitative
responses to the questions asked in our post-evaluation questionnaire.

\smallskip
\noindent\emph{What information did you find most helpful when determining
what caused tests to fail?} Overall, 21~participants explicitly mentioned
some aspect of Holmes as being most helpful. For 6 of these participants,
all the information provided by Holmes was most helpful for cause
identification. Another 8 participants noted that specifically the similar
passing and failing tests were most helpful. For example, P36 stated these
similar tests when presented ``side by side'' made it ``easy to catch a bug.''

The other 6 participants stated the execution traces were most helpful. One
participant's response said that the parts of Holmes output that were most
helpful was the output ``showing method calls, parameters, and return
values.'' This was particularly true when there were multiple method calls in
an execution according to P26: ``it was useful to see what was being passed
to them and what they were returning.''

\smallskip
\noindent\emph{What information did you find most helpful when deciding
changes to make to the code?} Overall, 14 participants mentioned some aspect
of Holmes as being most helpful. Of these, 5 explicitly stated that the
similar passing tests were most helpful of the information provided by
Holmes. P7, who often manually modified failing tests to better understand
expected behavior noted ``it helped to see what tests were passing,'' which
helped him ``see what was actually expected and valid.''

For the other 4~participants, the execution traces were most helpful for
resolving the defect. One participant specifically mentioned that the return
values in the execution traces for passing and failing inputs were most
helpful because then he could tell ``which parts are wrong.''

\smallskip
\noindent\emph{Would you like to add any additional feedback to supplement
your responses?} Many participants used this question as an opportunity to
share why they thought Holmes was useful. Many reported comments such as
``Holmes is great!'' and ``really helpful.'' For many, Holmes was most useful
because it provided concrete, working examples of expected and non-expected
behavior that help with ``pinpointing the cause of the bug.''

A participant noted that without Holmes, they felt like it was ``a bit slower
to find the reason why the test failed.'' Another participant noted that the
trace provided by Holmes was ``somewhat more useful'' than the trace provided
by JUnit.

In free-form, unprompted comments throughout the study, participants often
mentioned that the passing and failing tests and traces were useful for their
tasks; several participants explicitly mentioned during their session that
having the additional passing and failing tests were ``super useful'' and
saved them time and effort in understanding and debugging the defect.

While the qualitative feedback is largely positive, it is important to point
out that we do not view Causal Testing tools as a replacement for JUnit. The
intent is for them to complement each other and help developers understand
and debug software behavior. Three participants explicitly mentioned that
Holmes is most useful in conjunction with JUnit and other tools available in
the IDE. Several participants highlighted the complementary nature of these
tools. For example, P26 explained that though Holmes was ``very useful when
debugging the code,'' it is most useful with other debugging tools as ``it
does not provide all information.''

Finally, participants also suggests ways to improve Holmes. One participant
mentioned that Holmes should add the ability to click on the output and jump
to the related code in the IDE. Another suggested making the differences
between the passing and failing tests visibly more explicit. Three
participants explicitly suggested, rather than bolding the entire fuzzed
input, only bolding the parts that are different from the original failing
test. Our findings suggest that \textbf{Causal Testing is useful for both
cause identification and defect resolution, and is complementary to other
debugging tools.}

\section{Causal Testing Applicability to Real-World Defects}
\label{sec:ApplicabilityEvaluation}

To evaluate the usefulness and applicability of Causal Testing to real-world
defects, we conducted an evaluation on the Defects4J
benchmark~\cite{Just14defects4j}. Defects4J is a collection of reproducible
defects found in real-world, open-source Java software projects: Apache
Commons Lang, Apache Commons Math, Closure compiler, JFreeChart, and
Joda-Time. For each defect, Defects4J provides a buggy version and fixed
version of the source code, along with the developer-written test suites,
which include one or more tests that fail on the buggy version but pass on
the fixed version.

We manually examined 330 defects in four of the five projects in the
Defects4J benchmark and categorized them based on whether Causal Testing
would work and whether it would be useful in identifying the root cause of
the defect. We excluded Joda-Time from our analysis because of difficulty
reproducing the defects.\footnote{Some such difficulties have been documented
in the Joda-Time issue tracker:\\
\url{https://github.com/dlew/joda-time-android/issues/37}.}

\subsection{Evaluation Process}
\label{subsec:criteria}

To determine applicability of Causal Testing to defects in the Defects4J
benchmark, we first imported the buggy version and fixed version into
Eclipse. We then executed the developer-written test suites on the buggy
version to identify the target failing tests and the methods they tested.

Once we identified the target failing tests and methods under test, we ran
Holmes using the target failing tests. If Holmes ran and produced causal test
pairs, we ran InTrace to produce execution traces. Sometimes, Holmes was
unable to produce an output. In these cases, we attempted to evaluate if a
more mature version of Holmes could have produced an output. To do this, we
manually made small perturbations to the test inputs in an attempt to produce
reasonably similar passing tests. We made perturbations based on the type of
input and how a more mature Causal Testing tool would work. For example, if
the test input was a number, we made small changes such as adding and
subtracting increments of one from the original value or making the number
positive or negative. We then executed the tests and attempted to produce
causal test pairs.

In cases where Holmes or our manual analysis was able to produce similar
passing tests, we next determined if this information could be useful for
understanding the root cause of that defect. To do this, we first used the
fixed version to determine what we believed to be the root cause. If we were
able to determine the root cause, we then made a determination on whether the
similar passing tests and execution information would help developers
understand the root cause and repair the defect.

We used this process and the produced information to categorize the defects,
as we describe next.

\subsection{Defect Applicability Categories}
\label{subsec:categories}

We categorized Causal Testing's applicability to each defect into the
following five categories:

\begin{enumerate}[label=\Roman*.]
	
\item \textbf{Works, useful, and fast.} For these defects, Causal Testing can
produce at least one minimally-different passing test that captures its root
cause. We reason Causal Testing would be helpful to developers. In our
estimate, the difference between the failing and minimally-different passing
tests is reasonably small that it can be found on a reasonable personal
computer, reasonably fast. For most of these defects, our existing Holmes
implementation was able to produce the useful output.

\item \textbf{Works, useful, but slow.} For these defects, Causal Testing can
produce at least one minimally-different passing test that captures its root
cause, and this would be helpful to developers. However, the difference
between the tests is large, and, in our estimation, Causal Testing would need
additional computation resources, such as running overnight or access to
cloud computing. For most of these defects, our current Holmes implementation
was unable to produce the necessary output, but a more mature version would.
	
\item \textbf{Works, but is not useful.} For these defects, Causal Testing
can produce at least one minimally different passing test, but in our
estimation, this test would not be useful to understanding the root cause of
the defect.
	
\item \textbf{Will not work.} For these defects, Causal Testing would not be
able to perturb the tests, and would tell the developer it cannot help right
away.
	
\item \textbf{We could not make a determination.} Because the defects in our
study are from real-world projects, some required project-specific domain
knowledge to understand. As we are not the original projects' developers, for
these defects, the lack of domain-specific knowledge prevented us from
understanding what information would help developers understand the root
cause and debug, and we elected not to speculate. As such, we opted not to
make an estimation of whether Causal Testing would be helpful for these
defects.
	
\end{enumerate}

\subsection{Results}
\label{subsec:eval_results} 

\begin{figure}[t]
	\centering
	\small
	\begin{tabular}{lcccccc}
		\toprule
		& \multicolumn{5}{c}{\textbf{Applicability Category}}                              &                \\
		\textbf{Project} & \textbf{I}   & \textbf{II}  & \textbf{III} & \textbf{IV}  & \textbf{V}    & \textbf{Total} \\
		\midrule
		Math             & 14           & 15           & 11           & 20           & \phantom{0}46 & 106            \\
		Lang             & 11           & \phantom{0}6 & \phantom{0}3 & 14           & \phantom{0}31 & \phantom{0}65  \\
		Chart            & \phantom{0}2 & \phantom{0}4 & \phantom{0}1 & \phantom{0}1 & \phantom{0}18 & \phantom{0}26  \\
		Closure          & \phantom{0}2 & 22           & \phantom{0}8 & \phantom{0}5 & \phantom{0}96 & 133            \\
		\midrule
		\textbf{Total}   & 29           & 47           & 23           & 40           & 191           & 330            \\
		\bottomrule
	\end{tabular}
	\caption{Distribution of defects across five applicability categories described in Section~\ref{subsec:categories}. }
	\label{fig:app_results}
\end{figure}

Figure~\ref{fig:app_results} shows our defect classification results. Of the
330~defects, we could make a determination for 139. Of these, Causal Testing
would try to produce causal test pairs for 99~(71\%). For the remaining
40~(29\%), Causal Testing would simply say it cannot help and would not waste
the developer's time. Of these 99~defects, for 76~(77\%), Causal Testing can
produce information helpful in identifying the root cause. For 29~(29\%), a
simple local IDE-based tool would work, and for 47~(47\%), a tool would need
more substantial resources, such as running overnight or on the cloud. The
remaining 23~(23\%) would not benefit from Causal Testing. Our findings
suggest that \textbf{Causal Testing produces results for 71\% of real-world
defects, and for 77\% of those, it can help developers identify and
understand the root cause of the defect.}

\section{Discussion}
\label{sec:Discussion}

Our findings suggest that Causal Testing can be useful for understanding root
causes and debugging defects. This section discusses implications of our
findings, as well as threats to the validity of our studies and limitations
of our approach.

\textbf{Encapsulating causality in generated tests.} Our user study found
that having passing and failing tests that are similar to the original failin
test that exposed a defect are useful for understanding and debugging
software defects, though not all defects. Participants found the passing
tests that provided examples of expected behavior useful for understanding
why a test failed. This suggests that Causal Testing can be used to generate
tests that encapsulate \emph{causality} in understanding defective behavior,
and that an important aspect of debugging is being able to identify expected
behavior when software is behaving unexpectedly.

\textbf{Execution information for defect understanding \& repair.} Execution
traces can be useful for finding the location of a defect~\cite{Dallmeier05},
and understanding software behavior~\cite{Beschastnikh11fse,
Beschastnikh13icse, Beschastnikh15tse, Beschastnikh11osr,
Beschastnikh11tool-demo-fse, Ghezzi14, Krka14fse, Ohmann14ase}.
Our study has shown that such traces can also be useful for understanding
root causes of defects, and, in some cases, can highlight these root causes
explicitly. Participants in our study found comparing execution traces useful
for understanding why the test was failing and how the code should behave
differently for a fix. For some participants, the execution trace information
was the most useful of all information provided. These results support
further use of execution traces when conducting causal experiments.

\textbf{Causal Testing as a complementary testing technique.} Our findings
support Causal Testing as a complement to existing debugging tools, such as
JUnit. Understandably, participants sometimes found themselves needing
information that Holmes does not provide, especially once they understood the
root cause and needed to repair the defect. Our findings suggest that Causal
Testing is most useful for root cause identification. Still, a majority of the
participants in our study found Holmes useful for both cause identification
and defect repair, despite, on average, taking longer to resolve defects with
Holmes. We speculate that increased familiarity with Causal Testing would
improve developers' ability to use the right tool at the right time, improving
debugging efficiency, as supported by prior studies~\cite{Johnson16}.

\looseness-1
\textbf{Supporting developers with useful tools.} The goal of software
development tools is often to decrease developer effort, such that developers
will want to use that tool in practice. However, research suggests that the
first thing practitioners consider when deciding whether to use a given tool
is that tool's usefulness~\cite{Riemenschneider02}.
Our study shows that participants often took more time to debug when using
Holmes; however, despite this and other challenges developers encountered,
participants still generally found Holmes useful for both understanding and
debugging defects. This suggests that an important part of evaluating a tool
intended for developer use is whether the tool provides useful information in
comparison to, or in our case, along with, existing tools available for the
same problem.

\subsection{Threats to Validity}
\label{sec:Threats}

\textbf{External Validity.}
Our studies used Defects4J defects, a collection of curated, real-world
defects. Our use of this well-known and widely-used benchmark of real-world
defects aims to ensure our results generalize. We selected defects for the
user study randomly from those that worked with our current implementation of
Holmes and that required little or no prior project or domain knowledge, with
varying levels of difficulty. The applicability evaluation considered all
defects across four projects.

The user study used 37~participants, which is within range of higher data
confidence and is above average for similar user studies~\cite{Faulkner03,
Smith15, Barik16, Muslu15tse}. Our study also relied on participants with
different backgrounds and experience.

\looseness-1
\textbf{Internal Validity.}
Our user study participants were volunteers. This leads to the potential for
self-selection bias. We were able to recruit a diverse set of participants,
somewhat mitigating this threat.

\textbf{Construct Validity.} Part of our analysis of whether
Causal Testing would apply and be useful for debugging specific defects was
manual. This leads to the potential for researcher bias. We minimized this
threat by developing and following concrete, reproducible methodology and
criteria for usefulness.

The user study asked participants to understand and debug code they had not
written, which may not be representative of a sitation in which developers
are debugging code they are familiar with (but is representative of a common
scenario of developers  debugging others' code). We aimed to select
defects for the study that required little project and domain knowledge.
Additionally, we did not disclose the true purpose of the user study to the
subjects until after the end of each participant's full session.

\subsection{Limitations and Future Work}
\label{sec:approach limitations}

Causal Testing mutates tests' inputs while keeping the oracles constant
(recall Section~\ref{subsec:fuzzing}). This process makes an implicit
assumption that small perturbations of the inputs should not affect the
expected behavior, and, thus, if small perturbations do affect the behavior,
knowing this information is useful to the developer for understanding the
root cause of why the faulty behavior is taking place. This assumption is
common in many domains, such as testing autonomous cars~\cite{Tianand18} and
other machine-learning-based systems~\cite{Pei17}. However, it also leads
Causal Testing limitations. In particular, some changes to the inputs
\emph{do} affect expected behavior, and using the unmodified oracle will not
be valid in these cases. This can lead Causal Testing to generate pairs of
tests that do not capture causal information about the expected behavior
properly. For example, it could produce a test that passes but that uses the
wrong oracle and should, in fact, fail. It remains an open question whether
such tests would be helpful for understanding root causes. The causal test
pair still indicates what minimal input change can satisfy the oracle, which
might still be useful for developers to understand the root causes, even if
the passing test does not properly capture the expected behavior.

Future work could extend Causal Testing to include oracle mutation. A
fruitful line of research, when specifications, formal or informal, are
available, is to extract oracles from those specifications. For example,
Swami~\cite{Motwani19icse} can extract test oracles (and generate tests) from
structured, natural language specifications, and Toradacu~\cite{Goffi16},
Jdoctor~\cite{Blasi18}, and @tComment~\cite{Tan12} can do so from Javadoc
specifications. Behavioral domain constraints~\cite{Galhotra17fse,
Angell18demo-fse, Aggarwal18}, data constraints~\cite{Ernst01, Muslu15issta,
Muslu13ni-fse}, or temporal constraints~\cite{Dwyer99, Beschastnikh11fse,
Beschastnikh13icse, Beschastnikh15tse, Ohmann14ase} can also act as oracles
for the generated tests.

By fuzzing existing tests and focusing on test inputs that are similar to the
original failing test, Causal Testing attempts to mitigate the risk that the
tests' oracle will not apply. In a sense, a test's inputs must satisfy a set
of criteria for the oracle to remain valid, and by modifying the inputs only
slightly (as defined by static or dynamic behavior), our hope is that in
sufficiently many cases, these criteria will not be violated. Future work
could consider implementing oracle-aware fuzzing that modifies inputs while
specifically attempting to keep the oracle valid.

In some cases, it may not be possible to generate passing tests by generating
new tests. For example, code that never throws an exception cannot have a
test pass if that test's oracle expects the exception to be thrown. In such
cases, Causal Testing will not produce false positive results for the
developer, and will simply say no causal information could be produced.

Our studies have identified that Causal Testing is often, but not always,
helpful. Future work can examine properties of defects or tests for which
Causal Testing is more effective at producing causal information, and for
which that causal information is more helpful to developers. This information
can, in turn, be used to improve Causal Testing.

\section{Related Work}
\label{sec:related}

\looseness-1
The closest work to Causal Testing is BugEx~\cite{Rosler12}, which is
also inspired by counterfactual causality. Given a failing test, BugEx uses
runtime information, such as whether a branch is taken, to find passing and
failing tests that differ with respect to that piece of information.
Darwin~\cite{Qi12} targets regression failures and uses concrete and symbolic
execution to synthesize new tests such that each test differs in control
flow when executed on the buggy and the non-buggy version of the code. By
contrast, Causal Testing requires only a single version of the code, and only
a single failing test, and generates pairs of tests that differ minimally
either statically or dynamically (or both) to help developers understand the
root cause of the defect.

Delta debugging~\cite{Zeller99, Zeller02} aims to help developers understand
the cause of a set of failing tests. Given a failing test, the underlying
\texttt{ddmin} algorithm minimizes that test's input such that removing any
other piece of the test makes the test pass~\cite{Hildebrandt00}. Delta
debugging can also be applied to a set of test-breaking code changes to
minimize that set, although in that scenario, multiple subsets that cannot be
reduced further are possible because of interactions between code
changes~\cite{Sukkerd13icse-nier, Zeller02}. By contrast, Causal Testing does
not minimize an input or a set of changes, but rather produces \emph{other}
inputs (not necessarily smaller) that differ minimally but cause relevant
behavioral changes. The two techniques are likely complementary in helping
developers debug.

When applied to code changes, delta debugging requires a correct code version
and a set of changes that introduce a bug. Iterative delta debugging does not
need the correct version, using the version history to produce a correct
version~\cite{Artho11}. Again, Causal Testing is complementary, though future
work could extend Causal Testing to consider the development history to guide
fuzzing.

Fault localization (also known as automated debugging) is concerned with
locating the line or lines of code responsible for a failing
test~\cite{Agrawal95, Jones02, Wong10}. Spectral fault localization uses the
frequency with which each code line executes on failing and passing tests
cases to identify the suspicious lines~\cite{Jones02, Souza16}. When tests
(or failing tests) are not available, static code elements or data about the
process that created the software can be used to locate suspicious
lines~\cite{Menzies07, Menzies10}. Accounting for redundancy in test suites
can improve spectral fault localization precision~\cite{Hao05, Hao05a}. MIMIC
can also improve fault localization precision by synthesizing additional
passing and failing executions~\cite{Zuddas14}, and Apollo can do so by
generating tests to maximize path constraint similarity~\cite{Artzi10}.
Statistical causal inference uses observational data to improve fault
localization precision~\cite{Baah10, Baah11}. Importantly, while statistical
causal inference aims to infer causality, it does not apply the
manipulationist approach~\cite{Woodward05} that Causal Testing uses; as a
result, Causal Testing can make more powerful statements about the causal
relationships it discovers. Unfortunately, research has shown that giving
developers the ground truth fault location (even from state-of-the-art fault
localization techniques) does not improve the developers' ability to repair
defects~\cite{Parnin11}, likely because understanding defect causes requires
understanding more code than just the lines that need to be edited. By
contrast, Causal Testing discovers the changes to software inputs that
\emph{cause} the behavioral differences, and a controlled experiment has
shown promise that Causal Testing positively affects the developers' ability
to understand defect causes.

Mutation testing targets a different problem than Causal Testing, and the
approaches differ significantly. Mutation testing mutates the source code to
evaluate the quality of a test suite~\cite{Just14a, Just14b}. Causal Testing
does not mutate source code (it perturbs test inputs) and helps developers
identify root causes of defects, rather than improve test suites (although
it does generate new tests.) In a special case of Causal Testing, when the
defect being analyzed is in software whose input is a program (e.g., 
compiler), Causal Testing may rely on code mutation operators to perturb the
inputs.

Reproducing field failures~\cite{Jin12} is an important part of debugging
complementary to most of the above-described techniques, including Causal
Testing, which require a failing test case. Field failures often tell more
about software behavior than in-house testing~\cite{Wang17icst}.

Fuzz testing is the process of changing existing tests to generate more
tests~\cite{Godefroid07, Godefroid08} (though, in industry, fuzz testing is
often synonymous with automated test input generation). Fuzz testing has been
used most often to identify security vulnerabilities~\cite{Godefroid08,
Walls15issre}. Fuzzing can be white-box, relying on the source
code~\cite{Godefroid08} or black-box, relying only on the specification or
input schema~\cite{Jung08, Walls15issre}. Causal Testing uses fuzz testing
and improvements to fuzz testing research can directly benefit Causal Testing
by helping it to find similar test inputs that lead to different behavior.
Fuzzing can be used on complex inputs, such as programs~\cite{Holler12},
which is necessary to apply Causal Testing to software with such inputs (as
is the case for Closure, one of the subject programs we have
studied). Fuzz testing by itself does not provide the developer with
information to help understand defects' root causes, though the failing test
cases it generates can certainly serve as a starting point.

The central goal of automated test generation (e.g.,
EvoSuite~\cite{Fraser13}, and Randoop~\cite{Pacheco07a}) and test fuzzing is
finding new failing test cases. For example, combining fuzz testing, delta
debugging, and traditional testing can identify new defects, e.g., in SMT
solvers~\cite{Brummayer09}. Automated test generation and fuzzing typically
generate test inputs, which can serve as regression tests~\cite{Fraser13} or
require humans to write test oracles. Without such oracles, one cannot know
if the tests pass or fail. Recent work on automatically extracting test
oracles from code comments can help~\cite{Blasi18, Goffi16, Tan12}.
Differential testing can also produce oracles by comparing the executions of
the same inputs on multiple implementations of the same
specification~\cite{Brubaker14, Chen15, Evans07, Kastner17, Srivastava11,
Yang11a}. Identifying defects by producing failing tests is the precursor to
Causal Testing, which uses a failing test to help developers understand the
defects' root cause.

\section{Contributions}
\label{sec:Contributions}

We have presented Causal Testing, a novel method for identifying root causes
of software defects that supplements existing testing and debugging tools.
Causal Testing is applicable to 71\% of real-world defects in the Defects4J
benchmark, and for 77\% of those, it can help developers identify the root
cause of the defect. Developers using Holmes, a proof-of-concept
implementation of Causal Testing, were more likely to correctly identify root
causes than without Holmes (86\%~vs.~80\% of the time). Majority of
developers who used Holmes found it most useful when attempting to understand
why a test failed and in some cases how to repair the defect. Overall, Causal
Testing shows promise for improving the debugging process, especially when
used together with other debugging tools.

\begin{acks}

This work is supported by the National Science Foundation under grants
no.\
CCF-1453474, 
IIS-1453543, 
and CCF-1744471, 
and by Google and Oracle Labs.

\end{acks}

\balance

\bibliographystyle{ACM-Reference-Format}

\end{document}

%% file: preamble.tex
\usepackage{amsmath}
\usepackage{amsfonts}
\usepackage{amssymb}
\usepackage{amsthm}
\usepackage{balance}
\usepackage{graphicx}
\usepackage{xspace}
\usepackage{pifont}
\usepackage{multirow}
\usepackage{array}
\usepackage{booktabs}
\usepackage{cancel}
\usepackage{microtype}
\usepackage{enumitem}
\usepackage{color,colortbl}
\usepackage{xcolor}
\usepackage{dblfloatfix}
\usepackage{fixltx2e}
\usepackage{subfig}
\usepackage[font=small]{caption}
\usepackage{framed}

\usepackage{listings}

\definecolor{drkgreen}{rgb}{0.0, 0.5, 0.0}
\newcommand\lt[1]{{\lstinline+#1+}} 
\renewcommand\t[1]{{\lstinline+#1+}} 
\definecolor{dkgreen}{rgb}{0,0.5,0}
\definecolor{dkred}{rgb}{0.5,0,0}
\definecolor{gray}{rgb}{0.5,0.5,0.5}
\let\origthelstnumber\thelstnumber
\makeatletter
\newcommand*\Suppressnumber{%
  \lst@AddToHook{OnNewLine}{%
    \let\thelstnumber\relax%
     \advance\c@lstnumber-\@ne\relax%
    }%
}

\newcommand*\Reactivatenumber{%
  \lst@AddToHook{OnNewLine}{%
   \let\thelstnumber\origthelstnumber%
   \advance\c@lstnumber\@ne\relax}%
}

\definecolor{LightGray}{gray}{0.9}
\definecolor{Gray}{gray}{0.8}

\usepackage{tikz}
\newcommand*\circled[1]{\tikz[baseline=(char.base)]{
            \node[shape=circle,draw,inner sep=1pt] (char) {#1};}}

\usepackage{hyperref}
\hypersetup{%
  bookmarksdepth = {-2},
  pdftitle = {\pdftitle},
  pdfkeywords = {},
  pdfauthor = {\pdfauthors},
  pdfstartview={FitH},
  urlcolor=black,
  linkcolor=black,
  citecolor=black,
  colorlinks=true,
}